\documentclass[11pt]{article}

\PassOptionsToPackage{numbers}{natbib}

\usepackage[preprint]{neurips_2018}

\usepackage[utf8]{inputenc} 
\usepackage[T1]{fontenc}    
\usepackage{hyperref}       
\usepackage{url}            
\usepackage{booktabs}       
\usepackage{amsfonts}       
\usepackage{nicefrac}       
\usepackage{microtype}      

\usepackage{amsmath}
\usepackage{graphicx}
\usepackage{algorithm}
\usepackage{algorithmic}
\usepackage{bbm}
\usepackage{color}

\newcommand{\ignore}[1]{}

\title{On the Equivalence of Automatic and Symbolic Differentiation}

\author{\vspace{-0.5cm} {\color{white}.}\\ \vspace{1ex}
  {\bf S\"oren Laue}\\
  TU Kaiserslautern\\
  Germany\\\vspace{1ex}
  \texttt{laue@cs.uni-kl.de}
}

\begin{document}

\maketitle

\begin{abstract}
We show that reverse mode automatic differentiation and symbolic differentiation are equivalent in the sense that they both perform the same operations when computing derivatives. This is in stark contrast to the common claim that they are substantially different. The difference is often illustrated by claiming that symbolic differentiation suffers from ``expression swell'' whereas automatic differentiation does not. Here, we show that this statement is not true. ``Expression swell'' refers to the phenomenon of a much larger representation of the derivative as opposed to the representation of the original function. 
\end{abstract}

\section*{Common Myths}
When distinguishing automatic differentiation from symbolic differentiation one frequently encounters the following myths:
\begin{enumerate}
\item Symbolic differentiation leads to ``expression swell'', i.e., careless symbolic differentiation can easily produce exponentially large symbolic expressions. Hence, automatic differentiation is more efficient than symbolic differentiation.
\item Symbolic and automatic differentiation are very different, especially in the presence of control flow statements like \texttt{if} or \texttt{while} loops.
\item Symbolic differentiation faces the problem of converting a computer program into a single expression.
\item After computing a derivative using symbolic differentiation, which might lead to an exponential growth, one can reduce the size of the resulting expressions by using common subexpressions. However, this is has to be done very carefully and is not straightforward and the intermediate representation will still see an exponential growth.
\end{enumerate}

The purpose of this paper is to shed some light onto these issues and show, that they are not true. In fact, we will show that reverse mode automatic differentiation and symbolic differentiation are equivalent.

\section{Introduction}

Computing derivatives is a fundamental task in computer science, especially in optimization and machine learning. Most optimization schemes rely on derivative information when minimizing a function. However, computing derivatives by hand is error prone and can be a time consuming task, especially when the function to be differentiated is more complex. Hence, methods for automatically computing derivatives have been designed. The two major approaches that can be found in the literature are automatic differentiation and symbolic differentiation. Symbolic differentiation is basically what one knows from high school. Automatic differentiation refers to the fact, that one can compute derivatives even for computer programs that compute functions. Automatic differentiation is often also referred to as algorithmic differentiation. The literature and tools on automatic differentiation is extensive, see, e.g., \cite{Andersson12,Bischof02b,Bischof92,Bischof02,Bucker06,Gebremedhin05,Griewank96,Griewank08,Hascoet13}. Its popularity increased significantly over the last few years, especially in the area of machine learning/deep learning~\cite{tf, Baydin18,LaueMG2018,LaueMG2019,LaueMG2020,LeCun2018,pytorch,Merrienboer2018b,Merrienboer2018a,Fei2018} where it is necessary to compute gradients of loss functions of deep nets.

It is often claimed in the literature that automatic differentiation is \emph{not} symbolic differentiation. To tell them apart, the phenomenon of ``expression swell'' is used~\cite[page~3]{Griewank08}. However, the difference is not really explained and stays in the dark. On the other hand it is sometimes hard to put one tool into its right category, i.e., either automatic differentiation or symbolic differentiation. For instance, the Theano framework~\cite{theano} that has been widely used in the machine learning community is sometimes said to use automatic differentiation and sometimes said to use symbolic differentiation~\cite{Baydin18}. Here, we explain why it is hard to tell both approaches apart by showing that they are in fact equivalent. They both perform the same operations when computing derivatives. The only difference they have is the underlying data structure. Automatic differentiation operates on directed acyclic graphs (DAGs) whereas symbolic differentiation operates on expression trees or expression forests if one allows common subexpressions. However, when we allow common subexpressions in symbolic differentiation, then reverse mode automatic differentiation and symbolic differentiation  compute the same result. We will also show that the phenomenon of ``expression swell'' does \emph{not} originate from the differentiation process but from transforming a DAG into a tree when disallowing common subexpressions.

\section{Expression Representation}

\begin{figure}[t!]
\vspace{-0.3cm}
  \centering
  \begin{minipage}{0.32\textwidth}
    \centering
    \includegraphics[height=6cm]{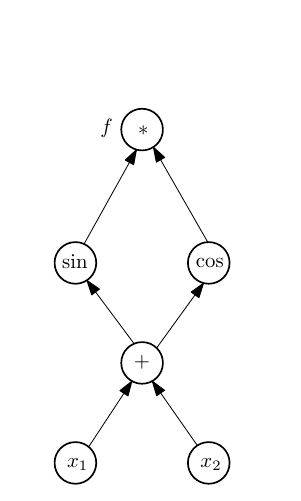}
\end{minipage}
  \begin{minipage}{0.32\textwidth}
    \centering
    \includegraphics[height=6cm]{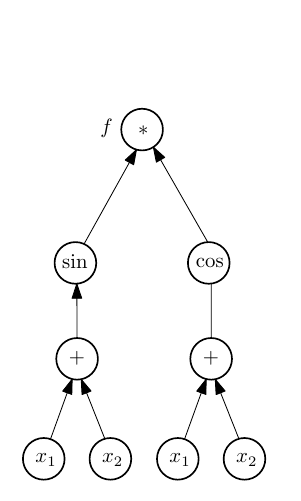}
  \end{minipage}
  \begin{minipage}{0.32\textwidth}
    \centering
    \includegraphics[height=6cm]{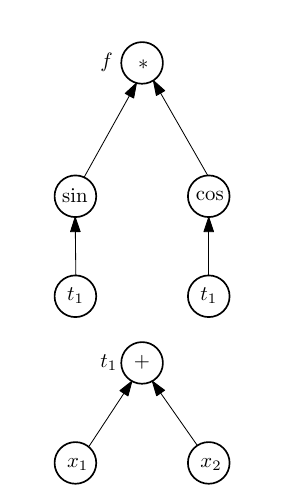}
  \end{minipage}
    \caption{Expression DAG (left), expression tree (middle), and expression forest with common subexpression $t_1=x_1+x_2$ (right) for the function $f(x_1, x_2) = \sin(x_1+x_2)\cos(x_1+x_2)$.} 
  \label{fig:expression}
\end{figure}
By expressions we understand mathematical expressions like $\sin(x_1+x_2)\cos(x_1+x_2)$. They can be represented by expression DAGs (also known as computational graphs or execution trace), expression trees, or expression forests with common subexpressions. Figure~\ref{fig:expression} illustrates the difference for the function $f(x) = \sin(x_1+x_2)\cos(x_1+x_2)$.
Obviously, an expression tree is also an expression DAG. On the other hand, an expression DAG can be converted into a tree simply by unfolding. The execution of a computer program results in an expression DAG. 
For instance, the following Python code results in the above expression DAG.

\vspace{1ex}
\begin{minipage}{0.2\textwidth}
\hspace{1cm}
\end{minipage}
\begin{minipage}{0.78\textwidth}
\begin{verbatim}
def f(x1, x2):
    t1 = x1 + x2
    f = sin(t1) * cos(t1)
    return f
\end{verbatim}
\end{minipage}
\vspace{2ex}

\begin{figure}[t!]
  \centering
  \begin{minipage}{0.1\textwidth}
    \centering
    \includegraphics[height=4.5cm]{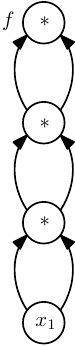}
\end{minipage}
  \begin{minipage}{0.58\textwidth}
    \centering
    \includegraphics[height=4.5cm]{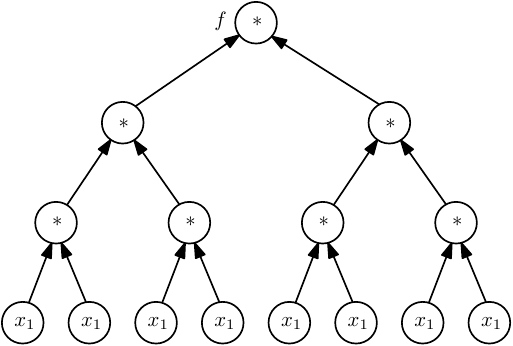}
  \end{minipage}
  \begin{minipage}{0.25\textwidth}
    \centering
    \includegraphics[height=4.5cm]{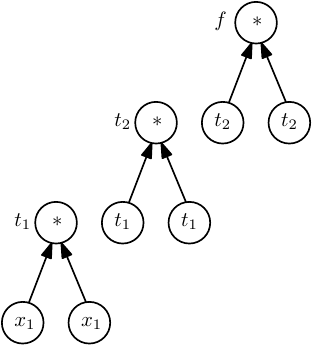}
  \end{minipage}
    \caption{Expression DAG (left), expression tree (middle), and expression forest (right) showing the phenomenon of ``expression swell'' and the use of common subexpressions to prevent it.} 
  \label{fig:exponential}
\end{figure}

When unfolding an expression DAG into an expression tree without common subexpressions it can become exponentially large. An example is given in Figure~\ref{fig:exponential}. However, when allowing common subexpressions it can be converted one-to-one where the resulting forest has the same size as the DAG. The following Python code corresponds to this example.

\vspace{1ex}
\begin{minipage}{0.2\textwidth}
\hspace{1cm}
\end{minipage}
\begin{minipage}{0.78\textwidth}
\begin{verbatim}
def f(x1):
    t1 = x1 * x1
    t2 = t1 * t1
    f = t2 * t2
    return f
\end{verbatim}
\end{minipage}
\vspace{2ex}

Now, we see an expression swell when converting a DAG into a tree. However, this is totally independent of computing derivatives. When allowing common subexpressions, we do not see an expression swell. This expression swell is often put forward as an argument for distinguishing between automatic and symbolic differentiation. Here, we see that it is \emph{not} connected to computing derivatives. Rather, it is a matter of the expression representation. Note, that when allowing common subexpressions such an expression swell does not occur.

\section{Equivalence of Reverse Mode Automatic Differentiation and Symbolic Differentiation}

In this section we will first review reverse mode automatic differentiation and symbolic differentiation, and then we will show their equivalence.

\subsection{Reverse Mode Automatic Differentiation}
Given an expression DAG $D=(V, E)$, reverse mode automatic differentiation proceeds as follows when computing the derivative of the output function $f$. Each internal node $v_i$ will eventually store the derivative $\frac{\partial f}{\partial v_i}$ that is commonly denoted as $\bar v_i$. Reverse mode proceeds from output to input nodes. At the nodes 
representing the output function $f$, the derivative $\frac{\partial f}{\partial f}$ is stored. Then, the derivatives that are
stored at the remaining nodes, here called $v_i$, are iteratively
computed by summing over all their outgoing edges using the following
equation:
\begin{equation} \label{eq:reverse}
\bar v_i = \frac{\partial f}{\partial v_i} = \sum_{v\,:\, (v_i, v) \in E} \frac{\partial f}{\partial v} \cdot \frac{\partial v}{\partial v_i} = \sum_{v\,:\, (v_i, v)\in E} \bar v \cdot \frac{\partial v_i}{\partial v},
\end{equation}
where the $\bar v = \frac{\partial f}{\partial v}$ are the partial derivatives that have been computed before and are stored at the nodes $v$. Finally, the derivative of the function $f$ with respect to all variables is stored at the corresponding input nodes.

\ignore{Let us illustrate reverse mode with the example $f(x_1, x_2) = \sin(x_1)\cos(x_1+x_2)$. The corresponding expression DAG can be found in Figure~\ref{fig:example}.}

\subsection{Symbolic Differentiation}
Symbolic differentiation applies the following two rules iteratively to a given function in order to compute its derivative. 

If the function is unary, e.g., sine, cosine, etc., it applies the following rule
\begin{equation} \label{eq:symbolic1}
\frac{\partial f(g(x))}{\partial x} = \frac{\partial f}{\partial g} \cdot \frac{\partial
  g}{\partial x}.
\end{equation}
In case of binary functions, e.g, addition, multiplication, etc., symbolic differentiation applies
\begin{equation} \label{eq:symbolic2}
\frac{\partial f(g_1(x), g_2(x))}{\partial x} = \frac{\partial f}{\partial g_1} \cdot \frac{\partial g_1}{\partial x} + \frac{\partial f}{\partial g_2}\cdot \frac{\partial g_2}{\partial x}.
\end{equation}

All rules from Calculus~101 can be reduced to these two rules. For instance, the multiplication rule is commonly known as $\frac{\partial (uv)}{\partial x} = u\frac{\partial v}{\partial x}+v\frac{\partial u}{\partial x}$. The following sequence shows that this follows from Equation~\eqref{eq:symbolic2}. We have
\begin{align*}
\frac{\partial (uv)}{\partial x} & = u\frac{\partial v}{\partial x} + v\frac{\partial u}{\partial x} \\
& = \frac{\partial (uv)}{\partial v}\cdot\frac{\partial v}{\partial x} + \frac{\partial (uv)}{\partial u}\cdot\frac{\partial u}{\partial x} \\
& = \frac{\partial f}{\partial g_1} \cdot \frac{\partial g_1}{\partial x} + \frac{\partial f}{\partial g_2}\cdot \frac{\partial g_2}{\partial x},
\end{align*}
where  the binary function $f(., .)$ is the multiplication operation, $g_1=u$, and $g_2=v$.

\ignore{
Let us illustrate the application of the differentiation rules~\eqref{eq:symbolic1} and \eqref{eq:symbolic2} by computing the derivative of $f(x_1, x_2) = \sin(x_1)\cos(x_1+x_2)$ with respect to $x_1$.
\begin{align*}
\frac{\partial f}{\partial x_1} & = \frac{\partial \sin(x_1)\cos(x_1+x_2)}{\partial x_1} \\
& = \cos(x_1+x_2) \cdot \frac{\partial \sin(x_1)}{\partial x_1} + \sin(x_1)\cdot \frac{\partial \cos(x_1+x_2)}{\partial x_1}\\
& = \cos(x_1+x_2) \cos(x_1) \cdot \frac{\partial x_1}{\partial x_1}+ \sin(x_1) (-\sin(x_1+x_2))\cdot\frac{\partial (x_1+x_2)}{\partial x_1}\\
& = \cos(x_1+x_2) \cos(x_1) \cdot 1 + \sin(x_1) (-\sin(x_1+x_2))\left(\frac{\partial (x_1+x_2)}{\partial x_1}\cdot\frac{\partial x_1}{\partial x_1}+\frac{\partial (x_1+x_2)}{\partial x_2}\cdot\frac{\partial x_2}{\partial x_1}\right)\\
& = \cos(x_1+x_2) \cos(x_1) \cdot 1 + \sin(x_1) (-\sin(x_1+x_2))\left(1\cdot\frac{\partial x_1}{\partial x_1}+1\cdot\frac{\partial x_2}{\partial x_1}\right)\\
& = \cos(x_1+x_2) \cos(x_1) - \sin(x_1)\sin(x_1+x_2)
\end{align*}
}
Please note, that the input expression is usually stored in an expression tree or expression forest with common subexpressions. Now there are two important questions that remain to be answered: 

\paragraph{Question~1:}
How does symbolic differentiation proceed when encountering common subexpressions? 

We illustrate this problem by the following example: Suppose we are differentiating a multiplication node $f=uv$. The derivative would be $\frac{\partial f}{\partial x} = u\frac{\partial v}{\partial x} + v\frac{\partial u}{\partial x}$. Note, that $u$ (and also $v$) appear in the original function as well as in the derivative. There are three possibilities to store the result. 
\begin{enumerate}
\item Copy the whole subtree representing $u$.
\item Store a pointer to the subtree representing $u$. 
\item Introduce $u$ as a common subexpression.
\end{enumerate}

All three options are valid possibilities. Option~1 will result in a tree as the output, Option~2 will result in a DAG, and Option~3 in a forest where each tree represents a common subexpression. 
Conceptually, Option~2 and Option~3 are identical. And this is also how one would implement symbolic differentiation. It is much easier to just store a pointer to the subtree $u$ instead of copying the whole subtree $u$. And since common subexpressions are allowed, turning this into a forest with common subexpressions is also trivial. 

\paragraph{Question~2:}
Should intermediate results be stored? 

Consider for instance the expression $f = u + u$. Blindly computing the derivative would require to compute the derivative of $u$ twice. Whenever there is the need for computing the derivative of a subtree $u$, then one can simply check, if it has been done before. In this case, one does not compute it again but can reuse the old result. 

\subsection{Equivalence}

Each internal node in an expression DAG can have either one or two incoming edges. In case a node has one incoming edge, the differentiation Rule~\eqref{eq:reverse} for reverse mode automatic differentiation is equivalent to the differentiation Rule~\eqref{eq:symbolic1} for symbolic differentiation. In case of two incoming edges, it is equivalent to rule~\eqref{eq:symbolic2} in symbolic differentiation. So obviously, both approaches perform the same operations.

In symbolic differentiation, if one stores only a pointer to common subtrees, or introduces a common subexpression (Option~2 and Option~3 in Question~1) then this is exactly how automatic differentiation proceeds.

Furthermore, if one stores intermediate results and reuses them (Question~2) then symbolic differentiation and reverse mode automatic differentiation are identical.

\section{Demystifying the Common Myths}
It should be obvious now, that the myths that are commonly encountered when talking about automatic and symbolic differentiation are not true. The distinction between them is rather artificial. Especially, the claim of the ``expression swell'', i.e., careless symbolic differentiation can easily produce exponentially large expressions is not true. Instead, it will lead to the same result as reverse mode automatic differentiation when intermediate results are stored as common subexpressions. This can be done easily. Hence, also the claim that common subexpressions need to be carefully determined \emph{after} the differentiation process is not true. They can be simply collected during the differentiation process.

When the input is a tree of size $n$, then the resulting tree of the derivative will be of size at most $O(n)$ when storing only pointers to common subtrees and at most $O(n^2)$ when copying every subtree during the symbolic differentiation process.  There is \emph{no} exponential growth here. This misconception can be explained maybe by the following fact. When turning a DAG into a tree, one might see an exponential growth. But this is unrelated to differentiation. And if one turns a DAG into a forest with common subexpression then the size of the output stays the same {\bf (Myth~1)}.

\subsection{Speelpenning's Example}
Often, Speelpenning's example~\cite{Speelpenning80} is brought forward to illustrate that symbolic differentiation is inefficient. Speelpenning's example is the following. Consider the function $f(x_1, x_2, \ldots, x_n)=x_1x_2\ldots x_n$. When computing the gradient with respect to $x=(x_1, x_2, \ldots, x_n)$, symbolic differentiation would output
\[
\nabla f = \begin{pmatrix}
    x_2x_3\ldots x_n \\
    x_1x_3\ldots x_n \\
    \vdots \\
    x_1x_2\ldots x_{n-1}
  \end{pmatrix}
\]
It is argued that the output is unnecessary large and there are quite a number of common subexpressions that now need to be identified. This is not true. What is displayed is the final output. However, this is the final output \emph{after} expression simplification and the removal of common subexpressions. In between, they have already been computed in the same way as in reverse mode automatic differentiation. If one looks at the individual steps when computing the derivatives symbolically, one can see that the result is exactly the same as reverse mode automatic differentiation. Hence, symbolic differentiation is as efficient as reverse mode automatic differentiation {\bf (Myth~4)}.

Summing up, Speelpenning's example rather serves as an example why reverse mode automatic differentiation is more efficient in case of many input variables and one output function. But the same holds true for standard symbolic differentiation.

\subsection{Control Structures}
When using operator overloading in automatic differentiation, control structures are easily circumvented since they do not appear in the computational graph, also known as the execution trace. That is, even when the code contains control structures, the computational graph/expression DAG will not. The same reasoning also applies to symbolic differentiation {\bf (Myth~2 and 3)}.
 
 \section{Conclusion}
We have shown in this note that reverse mode automatic differentiation and symbolic differentiation are algorithmically equivalent, i.e., they both perform the same set of operations when computing derivatives. The big contribution of automatic differentiation is rather of conceptual nature. It introduced the surprising idea that one can compute the derivative of any computer program. While this is a natural, yet very smart idea in hindsight, coming up with it for sure was not trivial. It has also led to many insights and algorithmic approaches like cross-country mode, edge elimination, etc., and to automatic differentiation tools that compute derivatives of even very complex computer programs.
 
\subsection*{Acknowledgments}
The author would like to thank Joachim Giesen, Torsten Bosse, and Martin Bücker for useful discussions on this topic. The work of Sören Laue is funded by Deutsche Forschungsgemeinschaft (DFG) under grant LA~2971/1-1.
 
\bibliographystyle{plain}
\bibliography{equivalence}
\end{document}